\documentclass[pre,superscriptaddress,twocolumn,a4paper,showpacs]{revtex4-1}
\usepackage{amssymb}
\usepackage{wasysym}
\usepackage{graphicx}
\usepackage{epsfig}
\usepackage[caption=false]{subfig}
\usepackage{float}
\usepackage[usenames, dvipsnames]{color}
\begin{document}

\title{The revival of the Baldwin Effect}

\author{Jos\'e F.  Fontanari}
\affiliation{Instituto de F\'{\i}sica de S\~ao Carlos,
  Universidade de S\~ao Paulo,
  Caixa Postal 369, 13560-970 S\~ao Carlos, S\~ao Paulo, Brazil}
  
  \author{Mauro Santos}
\affiliation{Departament de Gen\`{e}tica i de Microbiologia, Grup de Gen\`omica, Bioinform\`atica i  Biologia Evolutiva (GGBE),
Universitat Auton\`oma de Barcelona, 
08193 Bellaterra (Barcelona), Spain}

\begin{abstract}

The idea that a genetically fixed behavior evolved from the once differential learning ability of individuals that performed the behavior is known as the Baldwin effect. A highly influential paper [Hinton G.E., Nowlan S.J., 1987. How learning can guide evolution. Complex Syst. 1, 495--502] claimed that this effect can be observed {\it in silico}, but here we argue that what was actually shown is that the learning ability is easily selected for. Then we demonstrate the Baldwin effect to happen in the {\it in silico} scenario by estimating the probability and waiting times for the learned behavior to become innate. Depending on parameter values, we find that learning can increase the chance of fixation of the learned behavior by several orders of magnitude compared with the non-learning situation.

\end{abstract}

\maketitle

\section{Introduction}\label{sec:intro}

As pointed out by Maynard Smith, a recurrent issue in evolutionary biology is whether natural selection can explain the existence of complex structures that are of value to the organism only when fully formed \cite{Maynard_87}. This is a favorite topic of intelligent design creationists because it goes directly to the heart of Darwin's theory of evolution \cite{Darwin_59}: ``If it could be demonstrated that any complex organ existed, which could not possibly have been formed by numerous, successive, slight modifications, my theory would absolutely break down.'' Although Darwin could find out no such a case, intelligent design creationists coined the term `irreducible complexity' \cite{Behe_96} (Behe's focus is on adaptation at the molecular level) to argue that evolution cannot account for the intricate engineering found in all organisms: remove any part of a complex structure and the whole thing stops working. These `new' arguments were refuted by showing that a standard Darwinian process of descent with modification can explain the eventual complexity of adaptive molecular features \cite{Lynch_05}. On another level, a number of influential evolutionary psychologists and cognitive scientists claim that the human mind and the human language are examples of such complex structures, and do not feel comfortable with putative neo-Darwinian explanations \cite{Pinker_90,Dennet_91,Dennet_95,Briscoe_97,Deacon_97,Pinker_97,Calvin_00,Dor_01,Yamauchi_04}. These scholars do not obviously embrace the irreducible complexity argument of creationists. Instead, they assert that a much-neglected process during the Modern Synthesis in the 1930s and 1940s should be at work to explain the evolutionary emergence of such high-order phenotypic traits: Baldwinian evolution.

The idea of Baldwinian evolution has been available since the end of the 1800's and involves the notion that traits that were learned or acquired in earlier generations could be genetically fixed in the course of the evolution \cite{Baldwin_96,Morgan_96,Osborn_96}; the so-called Baldwin effect \cite{Simpson_53}. Because of the obvious Lamarckian scent and the absence of a Darwinian mechanism to implement the idea, the Baldwin effect remained on the fringe of evolutionary biology until 1987, when Hinton and Nowlan  offered a simple evolutionary algorithm that showed how learning could guide evolution \cite{Hinton_87} (see also \cite{Maynard_87,Dennet_03}). However, perhaps because of the computational limitations at the time, the original simulations as well as their numerous replications (e.g., \cite{Belew_90,Fontanari _90,Ackley_91,Harvey_93}) solely offered evidence that the inheritable flexible traits, which could be learned during the organism's lifetime, are selected. 

Here we run Hinton and Nowlan's  evolutionary algorithm until the finite-size population becomes genetically homogeneous and show that, for the original algorithm parameters \cite{Hinton_87}, learning increases the fixation probability of the target fixed trait by 6 orders of magnitude with respect to the non-learning situation, thus turning a virtual impossibility into a non-remarkable event. This is perhaps the essence of Baldwin effect as a theoretical explanation for non-reducible complex structures referred to above by evolutionary psychologists and cognitive scientists. It should be stressed, however, that we do not claim that these scholars are right; we simply demonstrate something they have taken for granted but that has not been proved in any of the numerous papers discussing Hinton and Nowlan's work.

The rest of this paper is organized as follows. In Section \ref{sec:model}, we describe Hinton and Nowlan's evolutionary algorithm and argue that following the short-time dynamics for a few runs is not enough to show the Baldwin effect. This effect is shown in Section \ref{sec:fix}, where we present the statistics of the fixation of the target fixed trait using a very large number of runs in which the dynamics is followed until the complete homogenization of the population. In Section \ref{sec:G}, we show that selection at the individual level does not optimize the learning parameter of Hinton and Nowlan's model. Finally, Section \ref{sec:disc} is reserved for our concluding remarks.

\section{Simulations of Hinton and Nowlan's model}\label{sec:model}

In their proof-of-concept  paper,  Hinton and Nowlan \cite{Hinton_87} proposed to test  the theoretical plausibility of the Baldwin effect   simulating the evolution of a population  of  $N$ haploid sexual  individuals, each represented by a chromosome of  $L$ loci with three alleles at each locus: $1$, $0$, and $?$.  It is assumed that the $L$ loci code for neural connections so that alleles $1$ specify innately correct connections, alleles $0$ innately incorrect connections, and alleles $?$ guessable (plastic) connections containing a switch that can be on (right) or off (wrong).
Learning consists of giving each individual up to a maximum of $G$ random combinations of switch settings (with equal probability for on and off) on every trial. Those individuals that have a $0$ allele at any locus will never produce the right connectivity of the neural network. On the other hand, if the combination of switch settings and the genetically specified connections produce the correct  neural network (i.e., a fully connected neural network) at trial  $g \leq G$  the individual stops guessing and becomes mature for mating.

To determine the mating probability, each individual $i=1,2, \ldots,N$  is evaluated according to  a fitness function $w_i$. In the case that individual $i$ has the $L$  alleles correctly set innately (i.e., it has the correct genotype),  it is assigned the maximum fitness value,  $w_i = L$. In the case that individual $i$ has $P$ correct alleles  and $Q = L- P$ plastic alleles,  its fitness is a random variable given by
\begin{equation}\label{wi}
w_i = 1 + \left ( L - 1\right) \left ( 1 - \frac{g}{G}  \right)
\end{equation}
if $g \leq G$ and $w_i = 1$, otherwise. Here the number of guesses $g =1,\ldots, \infty $  is a geometrically distributed random variable with success probability $1/2^Q$. Hence, even if the individual has learned the correct setting of  neural  switches, its fitness is lower than that of
an individual who was born with the correct setting. The difference is the cost of learning  $\gamma_g = g \left (L -1 \right) /G$ for $g \leq G$
and $\gamma_g =L-1$ otherwise.
Finally, in the case that individual $i$  has at least one innately incorrect allele, it is assigned the basal fitness value $w_i = 1$. 

The generations do not overlap and the population size is fixed, so that to create the next generation from the current one we must perform exactly $N$  matings. The parents in a mating are two different individuals that are chosen at random from the current generation with probability proportional to their fitness. The single offspring of each mating is generated by applying the one point crossover operation: we pick one point $ 1 \leq m \leq L-1$  at random from each of parents' chromosomes to form one offspring chromosome by taking all alleles from the first parent up to the crossover point $m$, and all alleles from the second parent beyond the crossover point.    Thus  the offspring will always be a recombinant chromosome. Of course, none of the learning is passed on to the offspring, which inherits the same allelic configuration that  their parents had at the different loci. 

In the absence of learning  ($G=1$ in Eq.\ (\ref{wi})), we have a `needle-in-the-haystack' problem for which  no search mechanisms can do much better than a random  search  on the configuration space.  However, provided the initial frequency $p_1$ of alleles 1   is not too low (say, $p_1 > 0.4$) and for the parameter setting $N=1000$ and $L=20$ used in Hinton and Nowlan simulations, the  evolutionary algorithm  can produce the correct genotype with a sporting chance and, somewhat surprising, once this individual is produced it rapidly takes over the population and reaches fixation, despite the disrupting effect of recombination  \cite {Santos_15}.

In a biological context, the  `needle-in-the-haystack' fitness landscape  entails the existence of structures that are useless until completely formed and  whose evolution  is difficult to explain  within a purely Darwinian perspective that, roughly speaking, views evolution as a hill-climbing process on a fitness landscape  \cite{Maynard_87}. 
Although such structures are unlikely to exist in nature -- we adhere to the standard view that any complex biological organ evolved from less complex but nonetheless useful (or with different function) organs -- Baldwin effect offers a theoretical framework to consider these landscapes within a purely Darwinian perspective. 
In Hinton and Nowlan's scenario, learning  creates an increased fitness zone around the `needle' by allowing individuals whose connections are near perfect to learn the correct setting \cite{Hinton_87,Belew_90}.

\begin{figure}[!ht]
  \begin{center}
\includegraphics[width=0.48\textwidth]{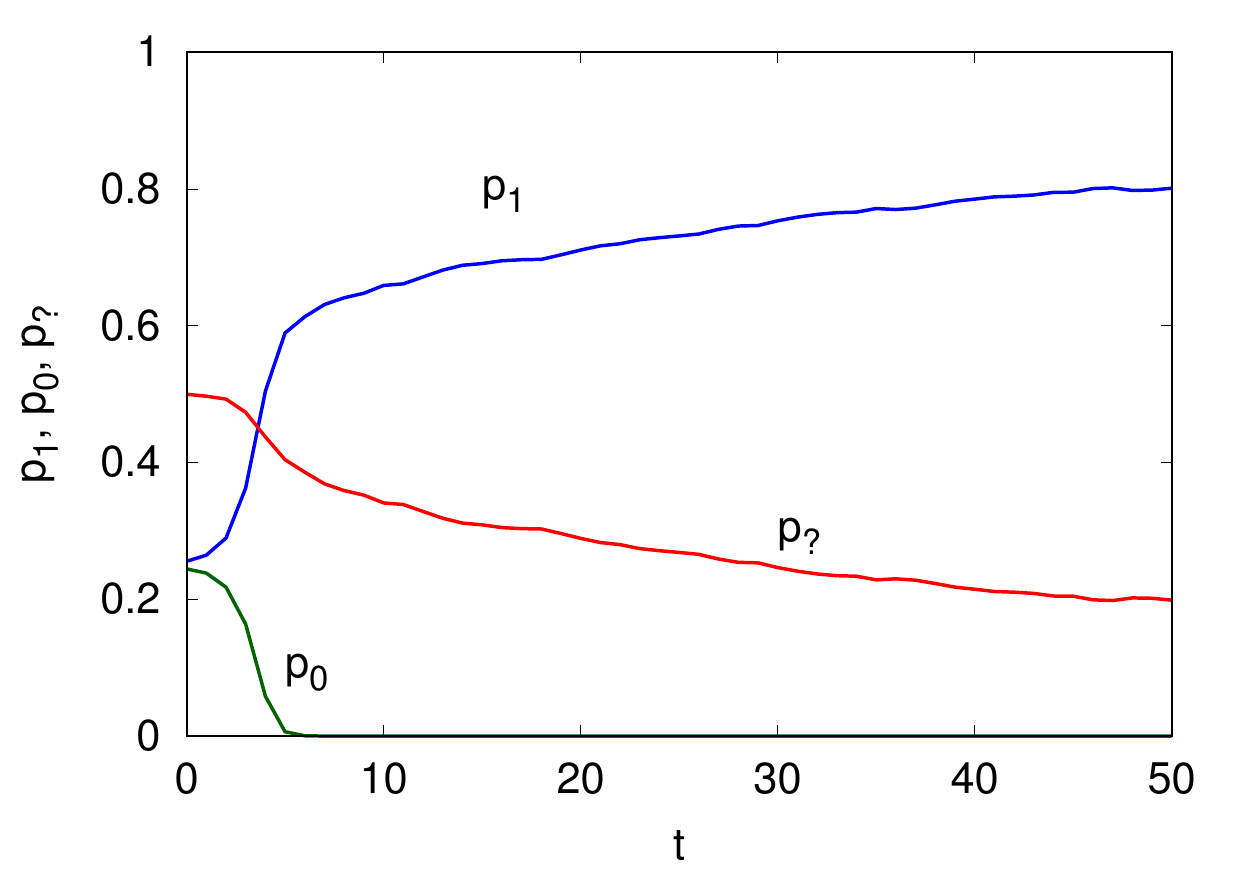}
  \end{center}
\caption{(Color online) Frequency of the three allele types,  $1$ (correct allele), $0$ (incorrect allele) and $?$ (plastic allele),   against  the number of generations  $t$  for a single run. This run does not lead to the  fixation of the correct genotype.  The parameters are $N=1000$, $L=20$ and $G=1000$.
 }
\label{fig:0}
\end{figure}

Somewhat surprisingly, although Baldwin's suggestion that behavioral goals in earlier generations could be genetically determined in the course of the evolution clearly points out to the necessity of looking at the fixation probability of the targeted traits, the quantitative studies around Hinton and Nowlan's work mainly focused on the time evolution of the allele frequencies. These studies did not analyze the fixation of allele $1$ in all the $L$ loci. For instance, Fig.\ \ref{fig:0} shows a typical result of a single run using the same parameters and time scale of the original simulations of Hinton and Nowlan (see Figure  2 of Ref.\ \cite{Hinton_87}). The population is able to quickly learn the solution as innately incorrect alleles $0$ are eliminated from the population, although the frequency of the plastic alleles $?$ remains relatively high. The trouble is that this type of graph shows only that the plastic alleles (and, consequently, learning) are selected, which is not exactly an extraordinary result. To demonstrate the Baldwin effect, the plastic alleles should be eliminated as well. For instance, in the particular run shown in  Fig.\ \ref{fig:0}, an allele $?$ fixed at one of the $L$ loci at generation $t=91$, thus precluding the fixation of the correct genotype. Undisputable evidence of Baldwin effect requires that we run the simulations until fixation of the correct genotype and average the results over very many runs. This is the aim of this paper.


\section{Fixation probability and mean time to fixation}\label{sec:fix}

 \begin{figure}
\centering
  	\subfloat {\includegraphics[width=0.48\textwidth]{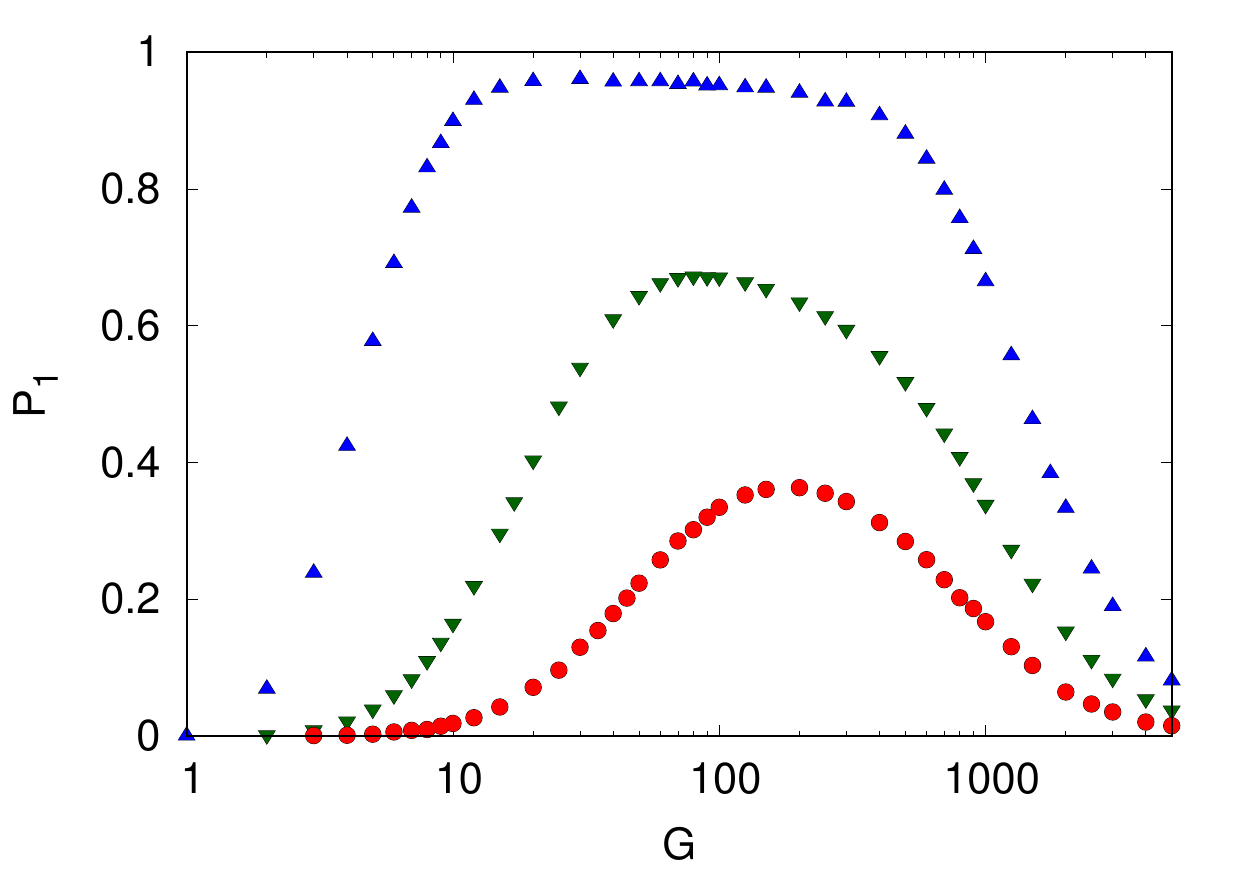}} \\
 	\subfloat {\includegraphics[width=0.48\textwidth]{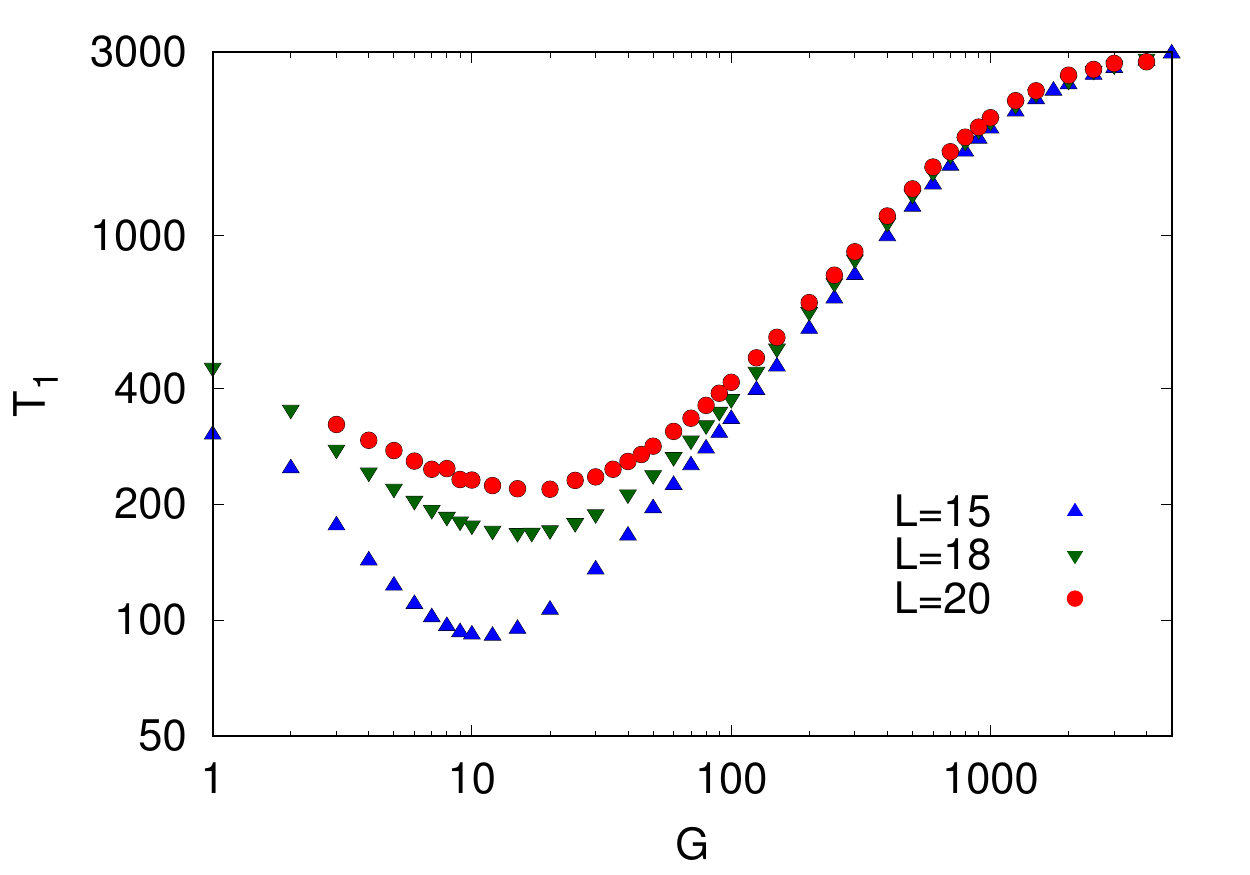}}
\caption{(Color online) Probability of fixation $P_1$ (upper panel)  and conditional mean time to fixation $T_1$ (lower panel) of the correct genotype  as functions of the maximum number of learning trials $G$ for $N=1000$  and three chromosome lengths $L=15$  ($\blacktriangle$), $L=18$  ($\blacktriangledown$) and $L=20$ ($\CIRCLE$) as indicated. For $L=20$ and $G=1000$ we find $P_1 = 0.167$ and $T_1 = 2026 $.
}
\label{fig:1}
\end{figure}

To calculate the  fixation probabilities  we must carry out a large number of independent simulations   and follow the dynamics until all $N$  chromosomes become identical (recall that in their original simulations Hinton and Nowlan neglected mutation). We denote by $P_1$  the fraction of runs in which we observed the fixation of
the correct genotype.  Note that since we are interested in $P_1$ only, we can abort a run whenever the fixation of  alleles $0$ or $?$ is observed at any loci. For instance, the run shown in Fig.\ \ref{fig:0} was aborted at generation $t=91$ when
the allele $?$ fixed at one of the $L$ loci. Hence, only the runs that lead to the fixation of the correct genotype must be followed  till the 
complete homogenization of the population.  

Another trick that greatly speeds up the simulation of the learning process consists of exploring the fact that number of guesses $g$ is a random variable distributed by a geometric distribution with probability of success $1/2^Q$, where $Q$ is the number of plastic alleles in the genome of the learner. In fact, we can easily obtain a geometric deviate $g \in \left \{1,2, \ldots \right \}$  from a uniform deviate  $r \in \left ( 0,1 \right )$  using the transformation method \cite{Press_92} 
\begin{equation}
g = \left  \lceil \frac{ \ln \left ( 1 - r \right )}{\ln \left ( 1 - 1/2^Q \right)} \right \rceil
\end{equation}
where the ceiling brackets notation   $\lceil  x \rceil $ stands for the  least integer greater than or equal to $x$.

The number of runs varies from $10^5$ to $10^8$ to guarantee that a statistically significant number of  fixations of the correct genotype  occurs. This is important because we use that sample  to estimate the
(conditional) mean time to fixation of the correct genotype, which we denote  by $T_1$.  Unless stated otherwise,  the initial frequency of alleles are set 
to $p_1=p_0 = 0.25$ and $p_? = 0.5$ in accord with Hinton and Nowlan's simulations. 

Figure  \ref{fig:1} summarizes our findings for the  population size $N=1000$. As the independent variable we choose  the parameter $G$, the reciprocal of which measures the difficulty  of learning. For instance, for large $G$ an individual with a few plastic alleles is certain to learn the correct  setting of switches and the learning cost $\gamma_g$ is typically  small. This results in  a quasi-neutral evolution scenario, where the fitness of the correct genotype differs very little from the fitness of genotypes with a few plastic alleles. 
 As expected, this  situation is  unfavorable to the fixation of the correct genotype and, accordingly, the upper panel of  Fig.\  \ref{fig:1} shows a drop of the fixation probability $P_1$ in the regime where learning is easy. For small $G$,  only individuals with a very small number of plastic alleles (and
none allele $0$) have a chance of guessing the correct switch setting.  For most individuals,  learning that setting is nearly impossible. The ineffectiveness of learning in this case is reflected in the very low probability of fixation of the correct genotype. In particular for $G=1$ (the non-learning limit), we find that $P_1 \approx 10^{-7}$ for $L=20$ (see the discussion  of Fig.\ \ref{fig:5} for the details of this estimate) whereas for 
$G=1000$ (the value chosen by Hinton and Nowlan) we find $P_1 \approx 0.167$. This gap of 6 orders of magnitude shows that Baldwin effect in Hinton and Nowlan's scenario  can make a virtual impossibility to happen with a sporting chance. Once the correct genotype fixed in the population,  there is no trace left of the plastic alleles, which acted as a scaffolding  mechanism to aid that fixation. The lower panel of Fig.\  \ref{fig:1} shows that the fixation takes place in a 
tenable time scale.

 \begin{figure}
\centering
  	\subfloat {\includegraphics[width=0.48\textwidth]{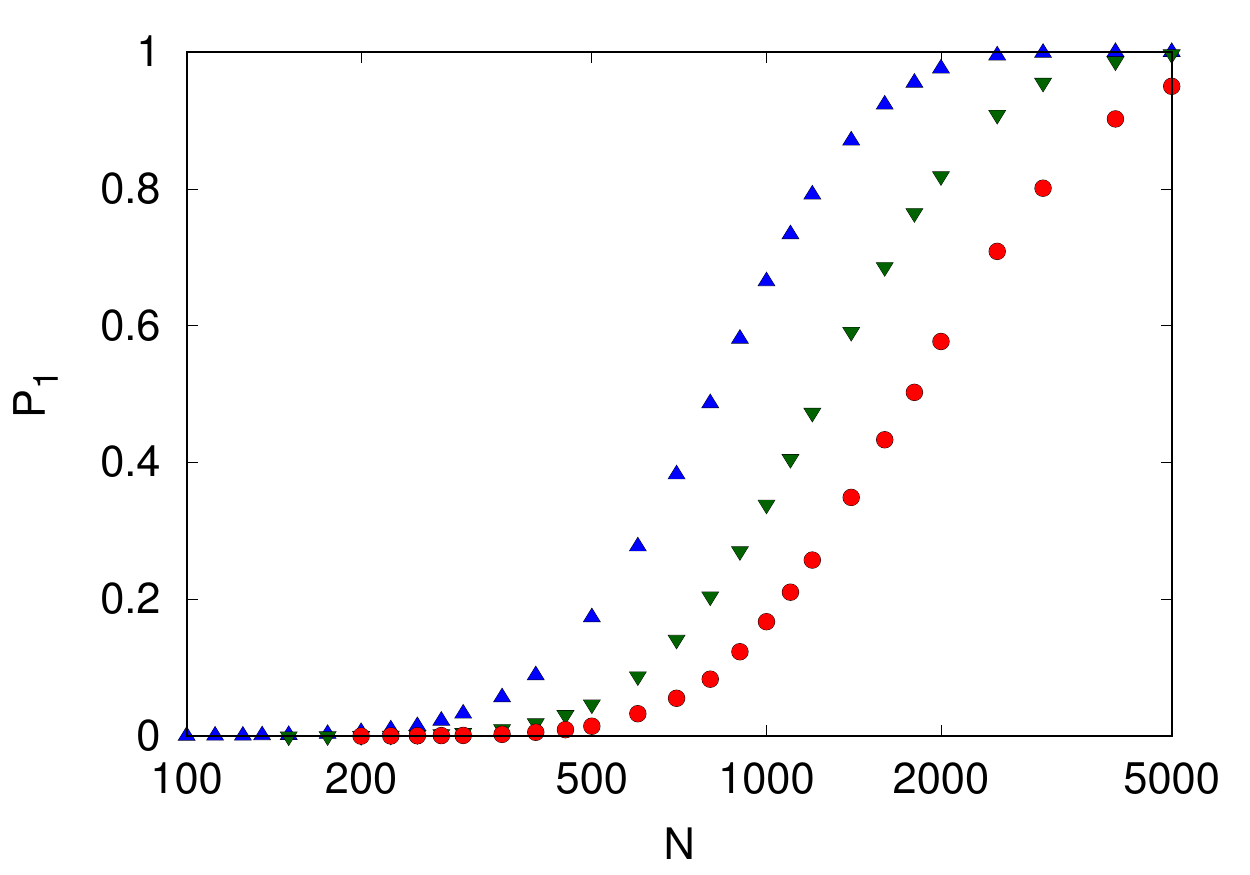}} \\
 	\subfloat {\includegraphics[width=0.48\textwidth]{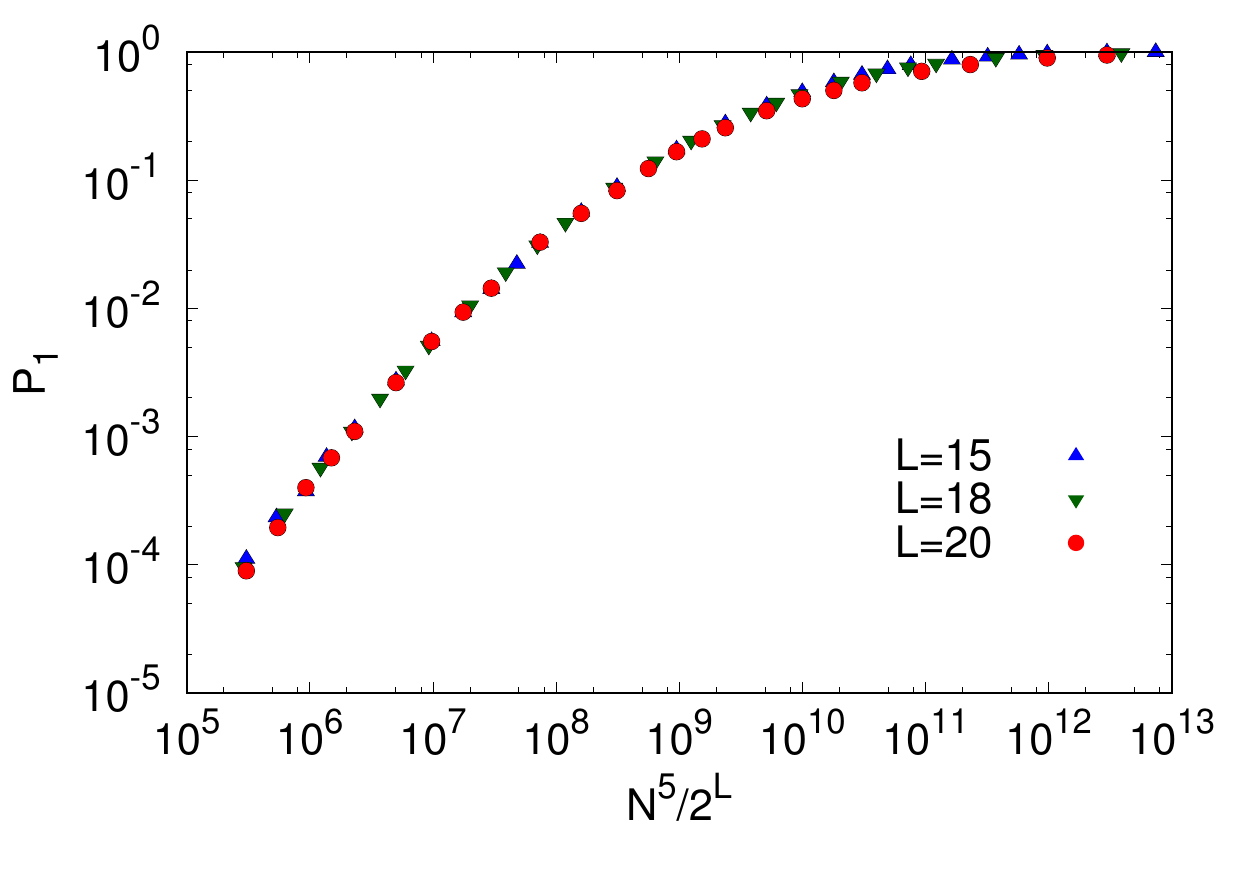}}
\caption{(Color online) Probability of fixation $P_1$ of the correct genotype  for $G=1000$  and  three chromosome lengths $L=15$  ($\blacktriangle$), $L=18$  ($\blacktriangledown$) and $L=20$ ($\CIRCLE$) as indicated. The  upper panel shows $P_1$ against the population size $N$ and the lower panel shows the collapse of the data for different $L$ when $P_1$ is plotted against the rescaled variable $N^5/2^L$.
}
\label{fig:2}
\end{figure}

Contrary to claims that the parameters of the original simulation of Hinton and Nowlan were very carefully selected \cite{Belew_90} to facilitate the `observation' of Baldwin effect,  the results shown in Fig.\ \ref{fig:1}  indicate that setting the maximum number of guesses to  $G=1000$  greatly  overestimate the values that optimize the fixation probability $P_1$ or the fixation time $T_1$. 
(We emphasize again   that the simulations of Hinton and Nowlan  offered no evidence of the Baldwin effect  -- they showed only that learning is selected.)  In particular, for the parameters of Fig.\ \ref{fig:1} we find that the optimal value of $G$ that maximizes the fixation probability $P_1$  is $G_{opt} \approx 2^{0.4 L}$. The exponential increase of $G_{opt}$ with  $L$ is expected since the number of switch settings to be explored  by the learning or guessing  procedure is $2^{p_? L}$, where $p_? = 0.5$ is the  frequency of the plastic allele in the initial population.

\begin{figure}[!ht]
  \begin{center}
\includegraphics[width=0.48\textwidth]{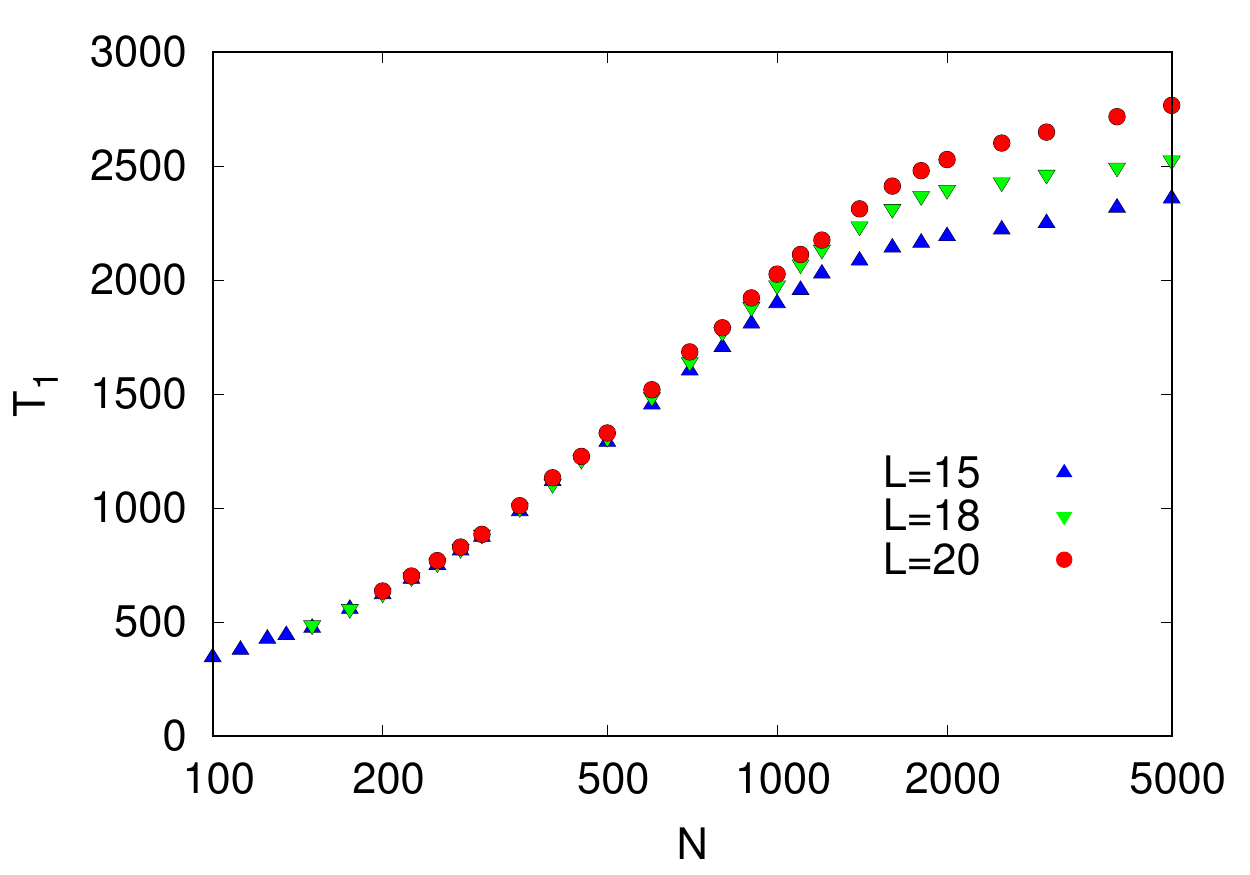}
  \end{center}
\caption{(Color online)  Conditional mean time to fixation $T_1$  of the correct genotype  as function of the population size $N$
 for $G=1000$  and three chromosome lengths $L=15$  ($\blacktriangle$), $L=18$  ($\blacktriangledown$) and $L=20$ ($\CIRCLE$) as indicated.
 }
\label{fig:2c}
\end{figure}

\begin{figure}[!ht]
  \begin{center}
\includegraphics[width=0.48\textwidth]{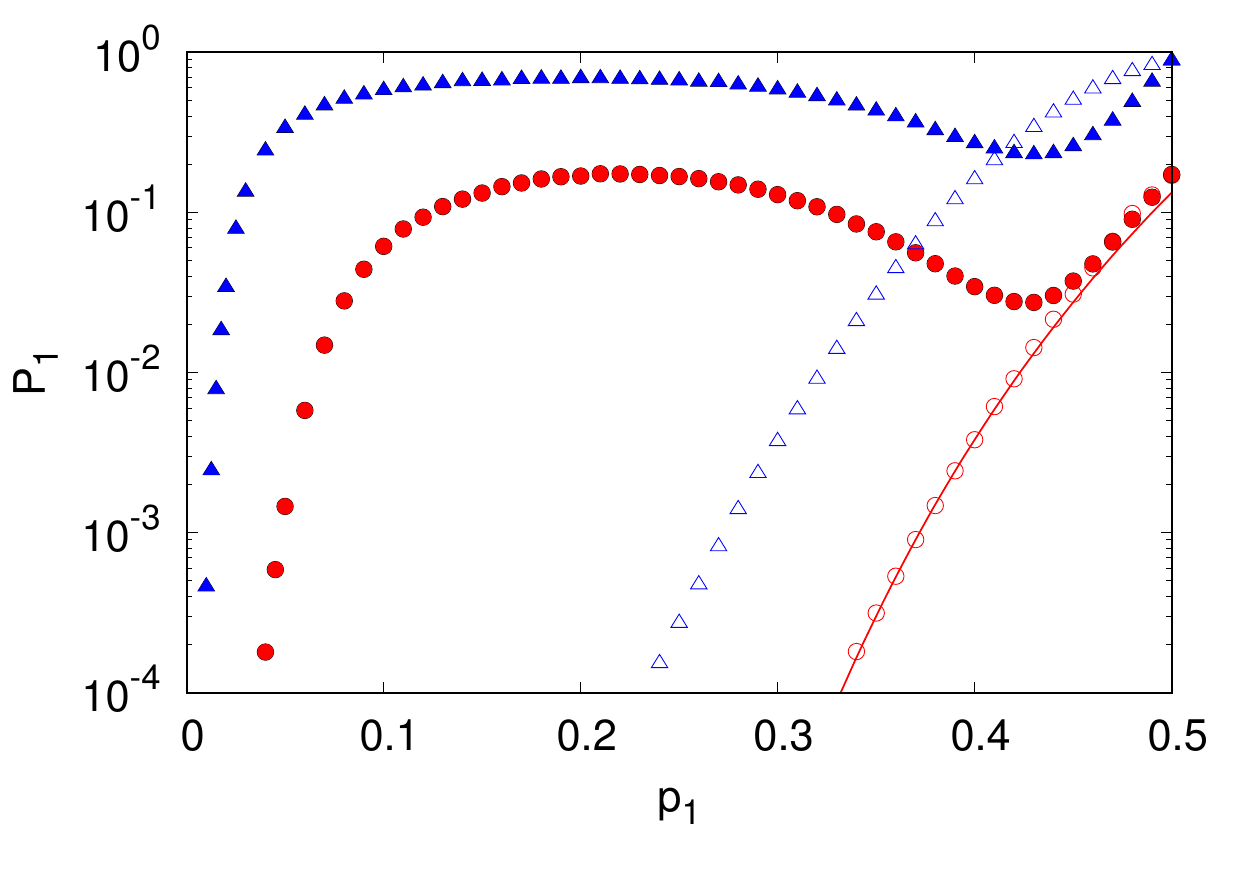}
  \end{center}
\caption{(Color online)  Probability of fixation $P_1$ of the correct genotype against the initial frequency of the correct allele $p_1$ for $N=1000$  and  two chromosome lengths $L=15$ ($\blacktriangle$, $\triangle$) and $L=20$ ($\CIRCLE$,$\circ$). The open symbols are the results for the non-learning case ($G=1$) whereas the filled symbols are the results for $G=1000$. The initial frequency of the incorrect allele is $p_0 = p_1$ and  the initial frequency of the plastic allele is $p_? = 1 - 2p_1$. The solid curve is the fitting of the data for $L=20$ and $G=1$ by the function $P_1 = 2^{a -b/p_1}$ with
$a \approx 17.6$ and $b \approx 10.2$.
 }
\label{fig:5}
\end{figure}

For fixed $G$ and $N$, increasing the chromosome length $L$ always results in the decrease of the probability of fixation (see Fig.\ \ref{fig:1}) and for large $L$ we find that $P_1$ decreases exponentially fast with increasing $L$.  This decrease can be compensated by increasing the population size $N$ as illustrated in Fig.\ \ref{fig:2}. In fact, in the regime where the fixation of the correct genotype is relatively rare, say $P_1 < 0.2$, this probability is practically unchanged with the increase of $L$ provided that $N$ increases such that the ratio $N^5/2^L$ is kept fixed.  In addition,  Fig.\ \ref{fig:2c} shows that the conditional mean time to fixation $T_1$ is a monotonically increasing function of $N$. In the region $N \ll 2^L$, we find $T_1 \propto \ln N$, regardless of the chromosome length $L$, whereas for large $N$ the fixation time  levels off towards asymptotic values that increase with increasing $L$.

The (theoretical) importance of the Baldwin effect is better appreciated when one considers the probability of fixation of the correct genotype $P_1$ as function of the initial frequency of the correct  allele $p_1$  in the population. In fact, if that allele is widespread in the initial population (say, $p_1 \approx 0.5$ for $L=20$ and $N=1000$) then the correct genotype has a sporting chance of fixation without need to invoke the Baldwinian evolution  \cite{Santos_15}.  Accordingly, in Fig.\ \ref{fig:5}  we show  $P_1$ against  $p_1$ for the non-learning case ($G=1$) and for the learning case  with  $G=1000$. The initial frequency of the incorrect  allele is set  to $p_0 = p_1$ and the initial frequency of the plastic allele is then $p_?= 1 - 2p_1$. Of course, for $p_1=0$ we have $P_1 =0$ since   the fixation of  the correct genotype is impossible if the correct allele is not present in the initial population.   Most interestingly, these results show that if the plastic allele  is rare (i.e., $p_1 \approx 0.5$)  then learning will  actually lessen the odds of fixation of the correct genotype when compared with the non-learning scenario. However,  when the plastic allele is  not under-represented in the initial population,   learning can produce an astronomical increase of those odds. For instance, since  in the range $p_1 > 0.34$ the data for $L=20$ and $G=1$ is very well fitted by the function $P_1 = 2^{a -b/p_1}$ with $a \approx 17.6$ and $b \approx 10.2$ (see  Fig.\ \ref{fig:5}) we can use this fitting to estimate the value of $P_1$ at $p_1=0.25$ and obtain $P_1 \approx 9 \times 10^{-8}$. Recalling that
$P_1 \approx 0.167$ for $G=1000$, we conclude that
the Baldwinian evolution boosts the chances of  fixation of the correct genotype by about 6 orders of magnitude for the parameter set used in the original simulation of Hinton and Nowlan. However, if those authors had used  $p_1 = 0.1$ instead, then that improvement would amount to 24 orders of magnitude.

\section{Adaptive maximum number of guesses}\label{sec:G}

We note that the value of the maximum number of guesses $G$ that maximizes the probability of fixation of the correct genotype (see Fig.\ \ref{fig:1})  cannot be selected by the evolutionary dynamics. To check that, instead of assuming that all individuals have
the same $G$,  we assign  uniformly distributed values of $G$ in the set $\{1,2, \ldots, 2000 \}$ to the $N$ individuals at generation $t=0$. More specifically, we introduce an additional locus to the chromosomes, say locus $L+1$, which stores the value of $G$. Similarly  to the other loci, this $G$-locus is inherited by the offspring.  In the rare event that the correct genotype appears in the randomly assembled initial population, we  assign the value $G=0$ to its $G$-locus. All the other chromosome types, regardless of their allelic compositions,  have their $G$-loci  assigned  to random values of $G$. 

\begin{figure}[!ht]
  \begin{center}
\includegraphics[width=0.48\textwidth]{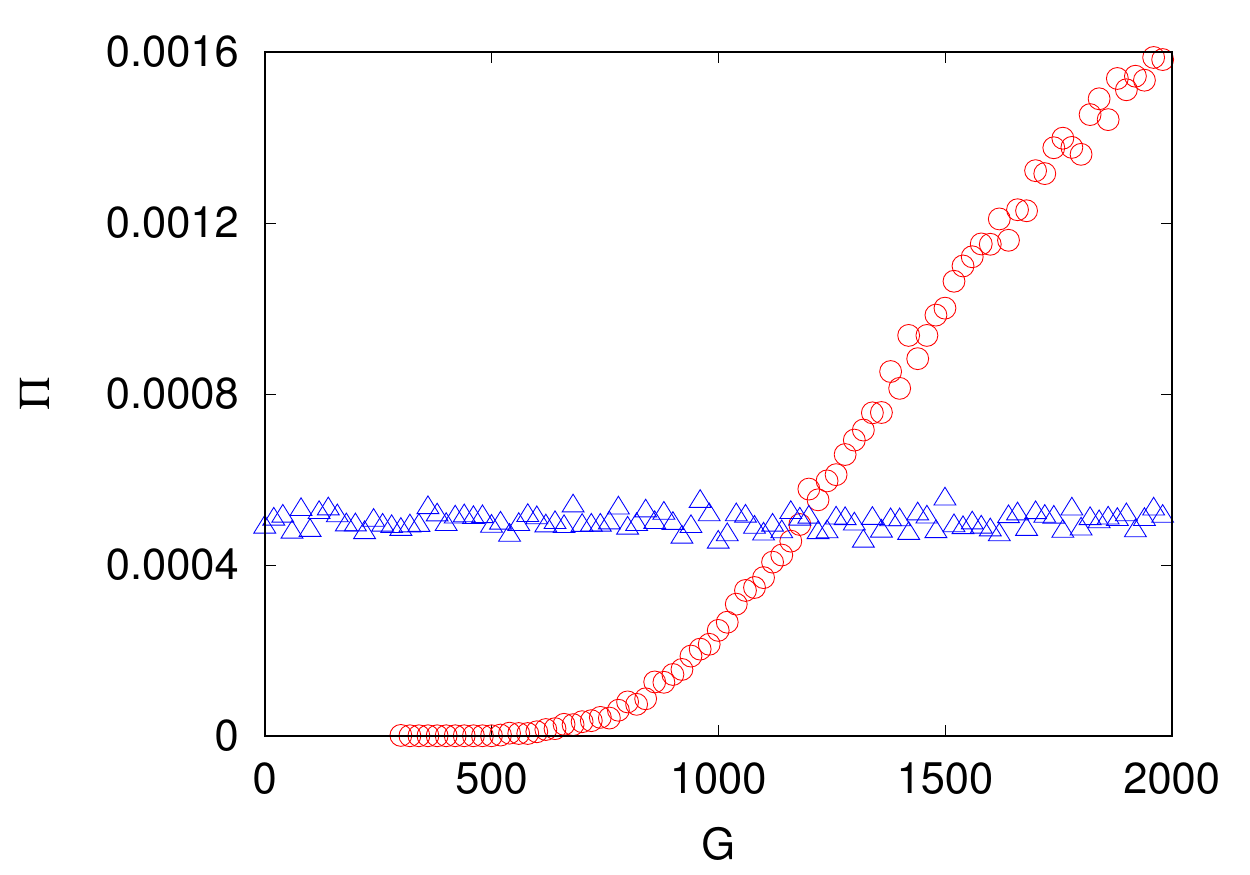}
  \end{center}
\caption{(Color online) Probability distribution $\Pi$ of the maximum number of guesses $G$ when the correct genotype fixed ($\circ$). The value of $G$ of each individual at generation $t=0$ is picked randomly from  the integers $\{1,2,\ldots,2000 \}$ and the symbols $\triangle$ show the initial distribution of $G$. The parameters are $N=1000$ and $L=20$.
 }
\label{fig:3}
\end{figure}

Figure \ref{fig:3} shows the probability distribution $\Pi$ of the values of $G$  for the runs that ended with the fixation of the correct genotype. For $10^5$ independent runs that led to the desired fixation, we found that $83\%$ of them resulted in the fixation of the $G$-locus as well (i.e., the individuals descended from a single ancestor), $16\%$ resulted in two distinct values of $G$ at the $G$-locus, $ 0.9\%$ in three values and $0.1\%$ in four values. The distribution of those values of $G$ is shown in Fig.\ \ref{fig:3} together with the initial uniform distribution. The results indicate that the fitness function in Eq.\  (\ref{wi})  favors individuals with large values of $G$. In fact, two individuals that used  the same number of guesses $g$ to find the correct switch setting will have different fitness  if they are assigned different $G$ values. All else being equal,   the fitness function in Eq.\  (\ref{wi})  increases with increasing $G$. 
Recalling that the maximum population fitness is achieved when the correct genotype fixed in the population,  we have here
an example where selection at the individual level, which  favors large $G$, conflicts with selection at the population level, which favors intermediate values of $G$. In fact, it would be interesting to see  what value of $G$  will dominate in a metapopulation scenario \cite{Fontanari_06}  since, in addition to the pressure of individual selection for the  largest possible $G$s, there is also a conflict
between  the value of $G$ that maximizes the fixation probability and  the value of $G$ that minimizes the 
fixation time of the correct genotype  (see Fig.\ \ref{fig:1}).

\section{Conclusions}\label{sec:disc}

In our previous work \cite{Santos_15} we criticized Hinton and Nowlan's paper \cite{Hinton_87} (and also Maynard Smith's \cite{Maynard_87}) on the grounds that they incorrectly assumed that a sexual population will never be able to fix the correct genotype, even if it is discovered many times, because recombination would always disrupt this genotype. By assuming that the frequency of alleles $1$ at  $t=0$  was $p_1 =0.5$, we proved Maynard Smith's claim that ``In a sexual population of 1000 with initial allele frequencies of 0.5, a fit individual would arise about once in 1000 generations \dots Mating would disrupt the optimum genotype, however, and its offspring would have lost the adaptation. In effect, a sexual population would never evolve the correct settings'' to be wrong because once the good genotype appears it is expected to increase exponentially and eventually fix. However, Fig.\ \ref{fig:1} and Fig.\ \ref{fig:5} show that if we set the initial frequency of alleles to $p_1 = p_0 = 0.25$  and $p_? = 0.5$   as in Hinton and Nowlan's simulations and do not allow learning, the good genotype has (as expected) very low chances of appearing in the population. 

An interesting result is that the number of queries (fitness evaluations) assumed by Hinton and Nowlan ($G \approx 2^{10}$  because at $t=0$  the average number of plastic alleles $?$ per individual is 10) does not maximize the fixation probability of the good genotype (Fig.\ \ref{fig:1}). The essence of the fitness function in  Eq.\  (\ref{wi}) is to alter the evolutionary dynamics by smoothing the fitness landscape and provide a direct path of increasing fitness to reach the highest peak. The problem is that the cost of learning, $\gamma_g = g \left ( L - 1 \right )/G$, decreases with increasing  $G$ which, in turn, flattens the shape of the fitness function and increases the `neutrality' zone around the peak. This makes the probability of fixation of the good genotype (Baldwin effect) highly dependent on the maximum number of guesses   allowed to each individual (Fig.\ \ref{fig:1}).

To sum up, our aim in this paper was to demonstrate and to quantify the probability of the Baldwin effect in the {\it in silico} scenario devised by Hinton and Nowlan \cite{Hinton_87}, something that has been surprisingly overlooked in the copious literature around this seminal work. Whether this effect offers a Darwin-compliant theoretical explanation to the evolution of non-reducible complex structures, or to the evolutionary emergence of high-order phenotypic traits such as consciousness or language, is ultimately an empirical question.

\acknowledgments
The research of JFF was  supported in part by grant
15/21689-2, Funda\c{c}\~ao de Amparo \`a Pesquisa do Estado de S\~ao Paulo 
(FAPESP) and by grant 303979/2013-5, Conselho Nacional de Desenvolvimento 
Cient\'{\i}\-fi\-co e Tecnol\'ogico (CNPq).  MS is funded by grant CGL2013-42432-P from the Ministerio de Econom\'{\i}a y Competitividad (Spain), and Grant 2014 SGR 1346 from Generalitat de Catalunya. This research used resources of the LCCA - Laboratory of Advanced Scientific Computation of the University of S\~ao Paulo.

\end{document}